\documentclass[aps,pre,reprint,groupedaddress]{revtex4-2}

\usepackage{amsmath}
\usepackage{amsfonts}
\usepackage{graphicx}
\usepackage{xcolor}
\usepackage{soul}
\usepackage{enumitem}

\begin{document}

\title{First-Passage Times for the Space-Fractional Spectral Fokker-Planck Equation}

\author{Christopher N. Angstmann}
\affiliation{%
	School of Mathematics and Statistics, University of New South Wales, Sydney NSW 2052 Australia
}%
\author{Daniel S. Han}
\email{daniel.han@unsw.edu.au}
\affiliation{%
	School of Mathematics and Statistics, University of New South Wales, Sydney NSW 2052 Australia
}%
\author{Bruce I. Henry}
\affiliation{%
	School of Mathematics and Statistics, University of New South Wales, Sydney NSW 2052 Australia
}%
\author{Boris Z. Huang}
\affiliation{%
	School of Mathematics and Statistics, University of New South Wales, Sydney NSW 2052 Australia
}%

\date{\today}

\begin{abstract}
We extend the random walk framework to include compounded steps, providing first-passage time (FPT) properties for a new class of superdiffusive processes, which are governed by the space-fractional spectral Fokker-Planck equation.
This first-passage process leads to novel FPT properties, different from L\'evy flights, that account for space dependent forces and hitting boundaries throughout the path of a jump.
The FPT distribution can be derived for different types of barriers and potentials, for which we also provide specific examples.
For the one-sided absorbing boundary with no potential on the semi-infinite line, we find that the FPT density scales asymptotically as $t^{-1/(2\alpha)-1}$ for large times, where the parameter $\alpha \in (0,1]$ relates to the power-law behavior for the distribution of the number of compounded steps. This is in agreement with the method of images but different to the Sparre-Andersen scaling $t^{-3/2}$ for corresponding L\'evy flights of order $2\alpha$.
In this case, there exists an optimal space-fractional exponent $\alpha$ to minimize the mean FPT.
\end{abstract}

\maketitle

\section{Introduction}
From molecular diffusion to financial markets, random walks are ubiquitously important in modeling many phenomena.
In such models, pivotal events may be triggered when a random walk first crosses a boundary or hits a target, conventionally called the first-passage time (FPT) or the first-hitting time \cite{redner2001guide}.
Problems involving FPTs are especially important in science, where collisions between particles determine chemical reaction rates for diffusing particles \cite{szabo1980first}, confined particles \cite{preston2021first}, particles following biased pathways \cite{park2003reaction}, time-dependent non-linear reaction networks \cite{rao2025exact}, cellular biology \cite{iyer2016first}, protein binding \cite{Reva2021} and in animal foraging strategies \cite{James2010,kilpatrick2021}, where the foraging pattern of animals dictates whether they find food and shelter \cite{Benichou2005,Benichou2011,viswanathan2008levy}.
The FPT remains a topic of increasing significance within numerous other recent studies on machine learning algorithms \cite{Sun2019,YANG2021}, trading strategies \cite{Jin2023} and quantum tunneling \cite{Zheltikov2023,Ni2023,Moosavi2024}.

The FPT of random walks exhibiting superdiffusive characteristics, where the mean square displacement (MSD) scales as $\langle x^2(t) \rangle \sim t^\nu$ for $\nu > 1$, has been of recent interest in theoretical studies.
Examples include: the optimization of searching for sparse targets \cite{Palyulin2014}, searches with external biases \cite{Palyulin_2014,padash2025first}, searches with subdiffusive waiting times \cite{koren2007first}, searches with short-lived targets \cite{boyer2024optimizing}, asymmetric searches \cite{padash2019first,padash2022asymmetric}, extreme statistics \cite{lawley2023extreme} and searches with restarts \cite{pal2017restart}.
In addition, there has also been vast applications within experimental settings such as in electronic relaxation in solutions \cite{mudra2020reaction}, interfacial diffusion \cite{wang2020non} and animal foraging strategies
\cite{viswanathan1999optimizing,viswanathan2008levy,viswanathan2011physics}.
The efficiency of superdiffusive models in the context of the animal foraging hypothesis has been the subject of sustained debate \cite{Benichou2011,Palyulin2014,levernier2020inverse,tzou2024counterexample,dipierro2023analysis}.
A well-established random walk model for superdiffusion is the L\'evy flight \cite{mandelbrot1979fractals}, where jump lengths have divergent variance and the trajectories are almost surely discontinuous.
These properties of L\'evy flights lead to difficulties in accounting for path-dependent effects such as barriers, boundaries and potential fields.
While L\'evy flights in the presence boundaries have been explored \cite{Zoia2007, Garbaczewski2019}, the inherent path discontinuity introduces two additional difficulties: particles can traverse potential barriers of arbitrary height in a single jump, destroying detailed balance and the Boltzmann steady state \cite{chechkin2003bifurcation,chechkin2004levy} and the non-locality of the Riesz operator makes the fractional Fokker–Planck equation analytically unwieldy in inhomogeneous or bounded domains.
Generalizations of such models have yielded the fractional Fokker-Planck equation by incorporating space-dependent advection forces \cite{metzler1999deriving,schertzer2001fractional} and asymmetric stable distributions \cite{dybiec2007stationary}.
The first-passage properties of these fractional Fokker-Planck equations have been studied in the context of escape from a ball \cite{blumenthal1961distribution,getoor1961first} and from potential wells \cite{chechkin2003bifurcation,chechkin2007barrier}.
While these processes account for the effect of potentials at the location of the random walker, the discontinuous paths still allow for the possibility of leaping over barriers without being subject to their effects.

Another well studied model for superdiffusion is the Levy Walk model in which Levy Flights occur with a finite velocity \cite{zaburdaev2015levy}.
In the simplest formulation, the points of visitation of the Levy Walk and the Levy Flight are identical, it is only the time of visitation of the points that is different \cite{palyulin2019first}.
Since the possibility of leaping over a target can still occur, the first passage time density and the first arrival time density are different for the Levy Flight model, as well as for the corresponding Levy Walk model \cite{palyulin2019first}.
Model equations and theoretical results for Levy Walk models with continuous trajectories have also been considered \cite{magdziarz2020limit} but the dynamical behavior of these models cannot be described using Fokker-Planck equations and is beyond the scope of this study.

A recent study has shown that superdiffusion and the space fractional Fokker-Planck equation can arise from compounded random walks \cite{Angstmann2025} with continuous trajectories.
These compounded random walks share the same governing equation as the L\'evy flight of order $2\alpha$ on $\mathbb{R}$ \cite{Metzler2000}: 
\begin{equation}\label{eq:FracDiffEq}
    \dfrac{\partial \rho (x,t)}{\partial t} = -D_{\alpha}(-\nabla^2)^\alpha \rho(x,t),
\end{equation}
where $\rho(x,t)$ is the probability density of finding the random walker at $x$, $\alpha\in(0,1]$ is the fractional exponent, $D_{\alpha}$ is the anomalous diffusion coefficient and $(-\nabla^2)^\alpha$ is the spectral fractional Laplacian, which can be defined on bounded or unbounded domains \cite{lischke2020fractional}.

In this paper, we explore the FPT of the compounded random walk \cite{Angstmann2025} and show that the FPT properties differ from L\'evy flights.
Physically, the differences arise due to the compounded random walk having connected paths in the continuum limit. 
Unlike previous formulations, this compounding allows the superdiffusive particle to interact with boundaries and sample local forces across its entire path. 
The link between the compounded random walk and the spectral fractional Laplacian \cite{Angstmann2025} allows us to formulate FPT problems in the spectral sense.
This is possible due to the homogeneous boundary conditions which arise from the formulation, as opposed to the exterior conditions necessary in L\'evy flight models.
In addition, the compounded random walk can be easily extended to include space-dependent biases, which leads to the space-fractional spectral Fokker-Planck equation \cite{Sokolov2001,Brockmann2002,Angstmann2025}
\begin{equation}\label{specfracFP}
    \dfrac{\partial \rho(x,t)}{\partial t} = -D_{\alpha}(-\mathcal{L})^\alpha \rho(x,t),
\end{equation}
where $(-\mathcal{L})^\alpha$ is the space-fractional spectral Fokker-Planck operator and $D_{\alpha}$ is the anomalous diffusion coefficient. 
This operator is spectrally defined by 
\begin{equation}
    (-\mathcal{L})^\alpha \rho(x,t) = \sum_{n=1}^\infty \langle\rho,\phi_n\rangle \lambda_n^\alpha \phi_n(x).
\end{equation}
Here, $\phi_n$ is the $n$\textsuperscript{th} normalized eigenfunction, with eigenvalue $\lambda_n$, and $\langle \cdot,\cdot \rangle$ is the weighted inner product associated with the Fokker-Planck operator $\mathcal{L}$ defined by
\begin{equation}
    \mathcal{L} \rho(x,t) = \dfrac{\partial}{\partial x}\left(\dfrac{\partial \rho(x,t)}{\partial x} + \beta V'(x)\rho(x,t)\right),
\end{equation}
with appropriate boundary conditions, where $\beta$ is a constant and $V(x)$ is the potential function. 
If $V(x)$ is constant, then \eqref{specfracFP} reduces to \eqref{eq:FracDiffEq} and, if $\alpha=1$, we recover the standard Fokker-Planck equation.

In what follows, we derive the FPT distribution from the compounded random walk, provide examples of FPT problems for different boundary conditions and outline a Monte Carlo scheme for stochastic simulations.
The examples on a finite interval and the semi-infintie line highlight how the FPT distribution, density and mean governed by \eqref{specfracFP} differ from the standard Fokker-Planck equation and L\'evy flight models.

\section{Model Framework}
Consider a discrete-time random walk on the set $\Omega = \{x_i = i\Delta x | 0 \leq i \leq M, \Delta x = L/M\}$, where a particle makes a step after every time increment. We denote the random variable $Y_n \in \Omega$ as the location of the particle after the $n$ time increments, given some initial position $Y_0 \in \Omega$. Let $\Omega' \subseteq \{0,L\} \subset \Omega$ be a set of points such that the particle is absorbed once it enters $\Omega'$. We are only concerned with the FPT into $\Omega'$ where it is absorbed, so the particle is considered to have stopped once it enters $\Omega'$. We will denote the probability of stepping from $x_i$ to $x_j$ via the probability transition function
\begin{equation}
	\Lambda_1 (x_i,x_j) = P(Y_n = x_j|Y_{n-1} = x_i),
\end{equation}
for all $n$. Thus, the probability to transition from $x_i$ to $x_j$ in exactly $n$ steps, without entering $\Omega'$ during the intermittent steps, is defined recursively by
\begin{equation}\label{nsteps}
	\Lambda_n(x_i,x_j) = \sum_{x_k \notin \Omega'} \Lambda_{n-1} (x_i,x_k) \Lambda_1(x_k,x_j).
\end{equation}
Note that the sum does not include the probability of stepping into the absorbing set so the sum over all $x_j$ may not be equal to $1$, which is to say
\begin{equation}\label{sum}
    \sum_{x_j \notin \Omega'} \Lambda_n(x_i,x_j) \leq \sum_{x_j \in \Omega} \Lambda_n(x_i,x_j) \leq 1.
\end{equation}
The FPT into $\Omega'$ is the first time step for which $Y_n \in \Omega'$ and is written as 
\begin{equation}
	n^* = \inf \{n|Y_n \in \Omega'\}.
\end{equation}
For a random walk initially at $x_i \notin \Omega'$, the probability of making $n$ steps without being absorbed is called the survival function given by
\begin{equation}
    P(n^* > n) = \sum_{x_j \notin \Omega'} \Lambda_n (x_i,x_j).
\end{equation}

\begin{figure}[h!]
	\includegraphics[width=\linewidth]{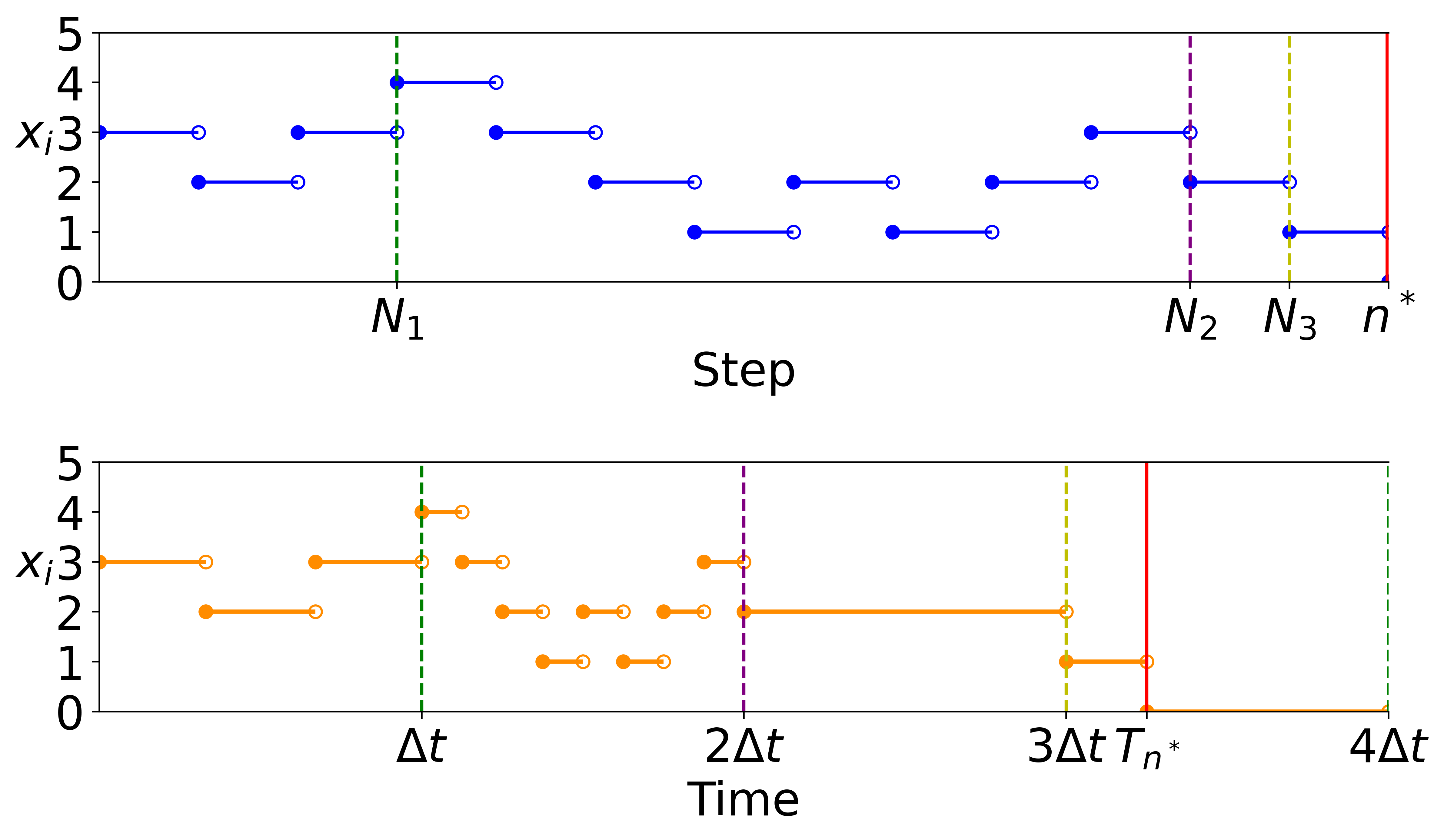}
	\caption{Example realization of a discrete time random walk, up to $n^*$ (top) where each step corresponds to a step in continuous time (bottom). Here, the domain is $\Omega = \{0,1,2,3,4,5\}$ with $\Omega' = \{0\}$ as the target region and $\Lambda_1(x_i,x_j) = 1/2$ for $|x_i-x_j| = 1$ or $x_i = x_j = 5$ and $\Lambda_1(x_i,x_j) = 0$ otherwise. The number of compounded steps in each interval are $K_1 = 3$, $K_2 = 8$, $K_3 = 1$ and $K_4 = 4$ but is absorbed before completing the $4$ steps between times $(3\Delta t,4\Delta t]$.}
	\label{fig:E}
\end{figure}

To obtain the FPT for the compounded random walk, we embed the steps from the discrete-time random walk above onto real values of time $t > 0$.
Suppose that for some $\Delta t>0$, the particle undergoes a compounded stepping process, where in the time interval $ ((m-1)\Delta t, m\Delta t]$, the particle steps $K_m$ times, where $K_m \geq 1$ is an i.i.d. random variable, or it has been absorbed into $\Omega'$.
The probability of making $k$ steps in the interval is denoted as $p(k) = P(K_m=k)$.
For convenience, we will assume that each of these $K_m$ steps are uniformly spaced in time over the interval with the last step occurring at time $m\Delta t$.
The total number of steps taken up to time $m\Delta t$ may be expressed as
\begin{equation}
	N_m = \sum_{i=1}^m K_i.
\end{equation}
We define $T_n \in [0,\infty)$ as the time when the $n$\textsuperscript{th} step occurs.
Then, the FPT of the compounded random walk is defined as the time when the $n^*$\textsuperscript{th} step occurs $T_{n^*} = \inf \{T_n|Y_n \in \Omega'\}$ and can be written as
\begin{equation}\label{eq:embedding}
    T_{n^*} = (m-1)\Delta t + \dfrac{n^*-N_{m-1}}{K_m}\Delta t,
\end{equation}
where $m = \sup \{i|n^* > N_{i-1}\}$,
and 
\begin{equation}
	(m-1)\Delta t < T_{n^*} \leq m\Delta t.
\end{equation}
A stochastic realization of the discrete time random walk, for $\Lambda_1(x_i,x_j) = 1/2$ for $|x_i-x_j| = 1$ or $x_i = x_j = 5$  and $\Lambda_1(x_i,x_j) = 0$ otherwise, and its embedding is illustrated in Figure \ref{fig:E}, where the dashed vertical lines show the corresponding pairs $(N_m,m\Delta t)$ and the solid red lines show the step and time when the random walk is absorbed into $\Omega'$.

To find the probability distribution of the FPT for the compounded random walk, we first denote the probability that the particle is at $x_j$ at time $m\Delta t$ by $\rho_\Delta (x_j,m\Delta t)$ where the subscript signifies the discrete space embedding.
Summing over all possible values of $K_m$ and $x_i$ using \eqref{nsteps}, we can write
\begin{equation}\label{gov}
    \rho_\Delta (x_j,m\Delta t) = \sum_{k=1}^\infty \sum_{x_i \notin \Omega'} \Lambda_k(x_i,x_j) p(k)\rho_\Delta (x_i,(m-1)\Delta t).
\end{equation}
Note that due to \eqref{sum},
\begin{equation}
    \sum_{x_j \notin \Omega'} \rho_\Delta (x_j,m\Delta t) \leq 1.
\end{equation}
To obtain the space-fractional Fokker-Planck equation \cite{Sokolov2001,Brockmann2002} from \eqref{gov}, we follow the approach in \cite{Angstmann2025} and use the Sibuya distribution \cite{sibuya1979generalized,sibuya1981generalized}
\begin{equation}\label{Sibuya}
    p(k) = \binom{\alpha}{k}(-1)^{k+1},
\end{equation}
for $k \in \mathbb{Z}^+$, with $0 < \alpha \leq 1$.
The distribution \eqref{Sibuya} implies the probability of a particle stopping at the $k$\textsuperscript{th} step given it has taken $k$ steps is $\alpha/k$ \cite{sibuya1979generalized,sibuya1981generalized}. 
The probability that $T_{n^*}$ is greater than the elapsed time $m\Delta t$ is then described by the survival function 
\begin{equation}\label{sur}
    \begin{split}
	\Psi_{n^*} (m\Delta t) &= P\left\{T_{n^*} > m\Delta t\right\}\\
    &= \sum_{x_j \notin \Omega'} \rho_\Delta (x_j,m\Delta t).
    \end{split}
\end{equation}
Survival up to time $m\Delta t$ implies that the particle has not entered into the target region $\Omega'$.

To derive the equation in continuous space, we embed $\Omega$ into the interval $[0,L] \in \mathbb{R}$.
We restrict the random walk to consist of biased nearest-neighbor steps via a Boltzmann weight \cite{henry2010fractional}.
The single-step probability transition function becomes \cite{Angstmann2025}
\begin{equation}
    \Lambda_1(x_i,x_j) = \dfrac{e^{-\beta V(x_j)}}{e^{-\beta V(x_i-\Delta x)}+e^{-\beta V(x_i+\Delta x)}},
\end{equation}
for $|x_i-x_j| = \Delta x$, where $V(x)$ is a potential and $\beta$ is the inverse temperature scale.
For $x_i \in \{0,M\}$, $\Lambda_1(x_i,x_i)$ is set depending on the desired boundary conditions, while for all other pairs of $x_i,x_j$, the transition probability $\Lambda_1(x_i,x_j) = 0$.
Now, we can now take a diffusion limit, where $\Delta x \rightarrow 0$ and $\Delta t \rightarrow 0$ such that
\begin{equation}\label{eq:D_alpha}
    D_\alpha = \lim\limits_{\substack{\Delta x \rightarrow 0\\ \Delta t \rightarrow 0}} \frac{\Delta x^{2\alpha}}{2^\alpha \Delta t},
\end{equation}
is the diffusion constant, $m\Delta t \rightarrow t$ and $x_i \rightarrow x$.
We let $\rho_\Delta (y_j,m\Delta t)/\Delta x \rightarrow \rho(x,t)$, where $\rho(x,t)$ is the probability density function (PDF) of finding the particle at $x$ at time $t$.
Therefore, modifying \eqref{gov} and taking the limit, we obtain \eqref{specfracFP}, the space-fractional spectral Fokker-Planck equation. 
Next, we take the limit of \eqref{sur} in the same way and obtain the survival function
\begin{equation}\label{survival}
    \Psi(t) = \int_0^L \rho(x,t) \ dx,
\end{equation}
where $\Psi_{n^*}(m\Delta t) \rightarrow \Psi (t)$ and $\rho(x,t)$ is the solution to \eqref{specfracFP}. 
We remark that the same governing equation and survival function is obtained in the limit for any choice of continuous time step embedding in \eqref{eq:embedding} as long as the distribution of the compounded steps obeys a power law $p(k) \sim k^{-1-\alpha}$.

The PDF of the FPT can be found via the relation $\psi(t) = -\Psi'(t)$.  The mean FPT is \cite{redner2001guide}
\begin{equation}\label{expectation}
		\langle T \rangle =  \int_0^\infty \Psi(t) \ dt,
\end{equation}
provided that $t\Psi(t) \rightarrow 0$ as $t \rightarrow \infty$.
Now, we examine the FPT and its properties for an interval with absorbing boundaries in one dimension.

\section{Superdiffusion with two absorbing boundaries}
First, we consider the case where the particle is between two absorbing boundaries at $x=0$ and $x=L$ with no potential $V(x) = 0$. The governing equation \eqref{specfracFP} reduces to \eqref{eq:FracDiffEq}, with boundary conditions $\rho(0,t)=\rho(L,t)=0$ and a Dirac delta initial condition $\rho(x,0)=\delta (x-x_0)$.
The solution of \eqref{eq:FracDiffEq} can be obtained via standard spectral methods to be
\begin{equation}\label{Unbiased2s_rho}
	\rho(x,t) = \dfrac{2}{L}\sum_{n=1}^\infty e^{-D_\alpha (n\pi/L)^{2\alpha} t} \sin \left(\dfrac{n\pi x_0}{L}\right) \sin \left(\dfrac{n\pi x}{L}\right),
\end{equation}
where $\rho(x,t)$ represents the PDF of finding the particle at $x$ for any given time $t$.
This simple form of the solution is due to the nature of the spectral Laplacian which scales each eigenfunction by its Laplacian eigenvalue to power of $\alpha$.
A notable property of \eqref{Unbiased2s_rho} is the decay rate of the first eigenvalue $(\pi/L)^{2\alpha}$, which dominates the asymptotic behavior of $\rho(x,t)$ for a particle beginning in the center.
If $L \leq \pi$, then this term of the sum decreases fastest when $\alpha = 1$.
However, for $L > \pi$, the first eigenmode will decay quicker for smaller values of $\alpha$.
\begin{figure}[h]
	\includegraphics[width=\linewidth]{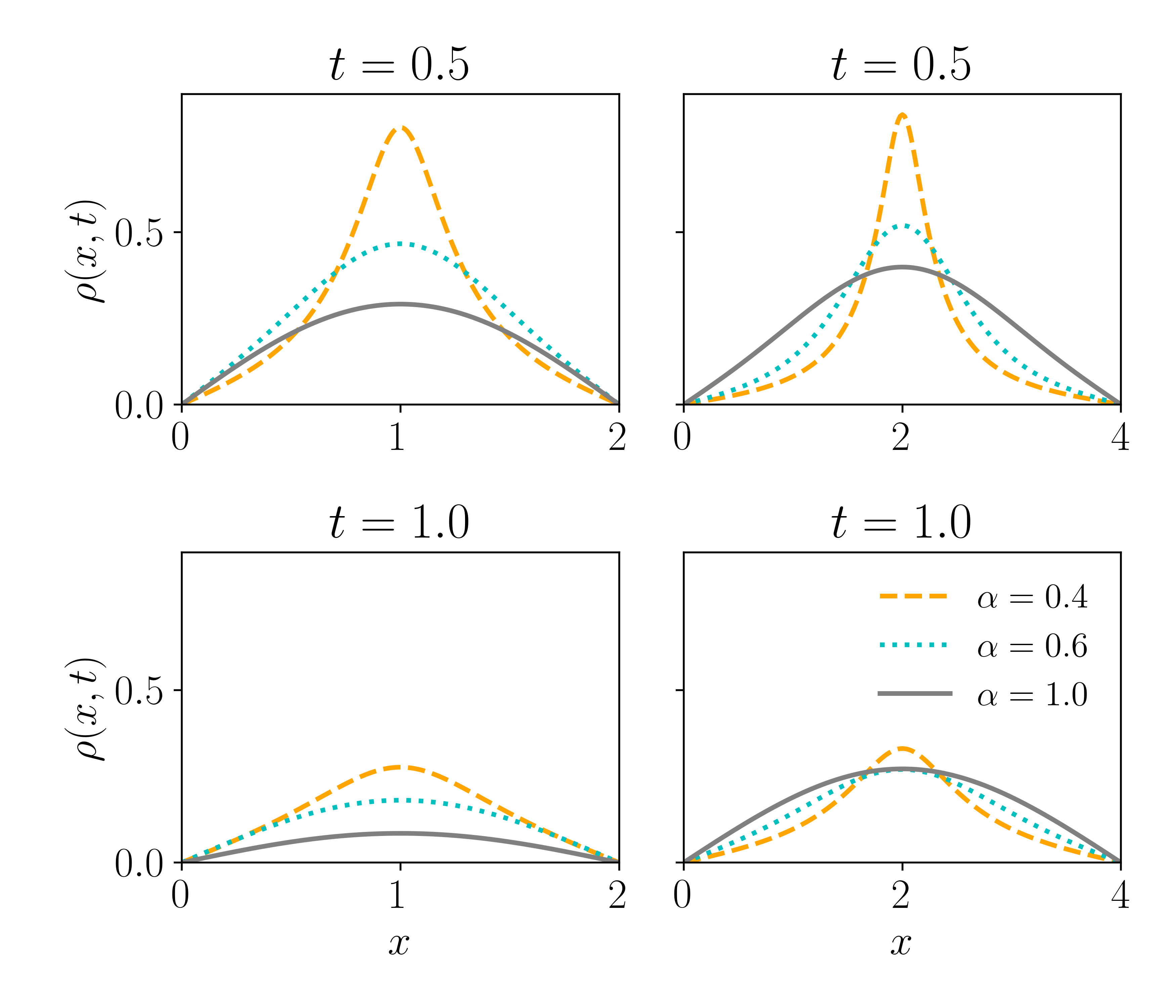}
	\caption{The PDF \eqref{Unbiased2s_rho} for finding the particle at $x$, where $0<x<L$ for $t = 0.5$ (top) and $1.0$ (bottom) truncated after the first $1000$ terms. Here, $D_{\alpha} = 1$, $L = 2$ (left) and $4$ (right), $x_0 = L/2$ for $\alpha = 0.4$, $0.6$ and $1.0$.}
	\label{fig:Unbiased2s_rho_c}
\end{figure}

This can be seen in Figure \ref{fig:Unbiased2s_rho_c}, which shows the PDF \eqref{Unbiased2s_rho} for various values of $\alpha$ over time for $L=2$ (Fig. \ref{fig:Unbiased2s_rho_c} left) and $L=4$ (Fig. \ref{fig:Unbiased2s_rho_c} right).
When $L>\pi$, the PDF \eqref{Unbiased2s_rho} decays slower for $\alpha = 1$ than when $L<\pi$ (Fig. \ref{fig:Unbiased2s_rho_c} left).
In contrast, when $\alpha$ is smaller ($\alpha=0.4$ or $0.6$), the rate of decay varies less as demonstrated by the difference between the peaks for $L=2$ (Fig. \ref{fig:Unbiased2s_rho_c}  left) and $L=4$ (Fig. \ref{fig:Unbiased2s_rho_c}  right).
Clearly, for small $\alpha$, $D_\alpha = \lim_{\substack{\Delta x \rightarrow 0\\\Delta t \rightarrow 0}} \Delta x^{2\alpha}/(2^\alpha \Delta t) = 1$ implies that the ratio between the step length and time increments $\Delta x/\Delta t$ approaches zero faster.
This means that the particle has a smaller step length and thus requires more steps to reach the boundary.
Eventually, it overcomes this by taking a large number of compounded steps, due to the divergent mean of the Sibuya distribution in each time increment $\Delta t$, such that the likelihood of absorption is increased.
If there is no absorbing region, then this process is superdiffusive with divergent MSD. However, we may characterize the spreading via the pseudo-MSD, defined as $\lim_{\delta \rightarrow 2\alpha} \langle |x|^{\delta} (t) \rangle$ for $0<\delta<2\alpha$, which scales as $t^{1/\alpha}$ \cite{Angstmann2025}. This is exactly the same as the L\'evy flight model \cite{Metzler2000}.

We can also see this in the survival function and PDF for the FPT.
Using equation \eqref{survival} we obtain the corresponding survival function
\begin{equation}\label{Unbiased2s_Sur}
	\begin{split}
		\Psi(t) &= 2\sum_{n=1}^\infty \dfrac{1-(-1)^n}{n\pi} e^{-D_\alpha (n\pi/L)^{2\alpha} t} \sin \left(\dfrac{n\pi x_0}{L} \right)\\
		&= 4\sum_{n=0}^\infty \dfrac{e^{-D_\alpha ((2n+1)\pi/L)^{2\alpha} t}}{(2n+1)\pi} \sin \left(\dfrac{(2n+1)\pi x_0}{L}\right),
	\end{split}
\end{equation}
and the PDF
\begin{equation}\label{Unbiased2s_PDF}
	\begin{split}
		\psi(t) &= \sum_{n=0}^\infty \dfrac{4D_\alpha e^{D_\alpha ((2n+1)\pi/L)^{2\alpha} t}}{((2n+1)\pi)^{1-2\alpha}L^{2\alpha}} \sin \left(\dfrac{(2n+1)\pi x_0}{L}\right).
	\end{split}
\end{equation}
Next, we consider the case for when the absorbing boundary on the right is sent to positive infinity.

\section{Superdiffusion with one absorbing boundary}

\begin{figure}[h!]
	\includegraphics[width=\linewidth]{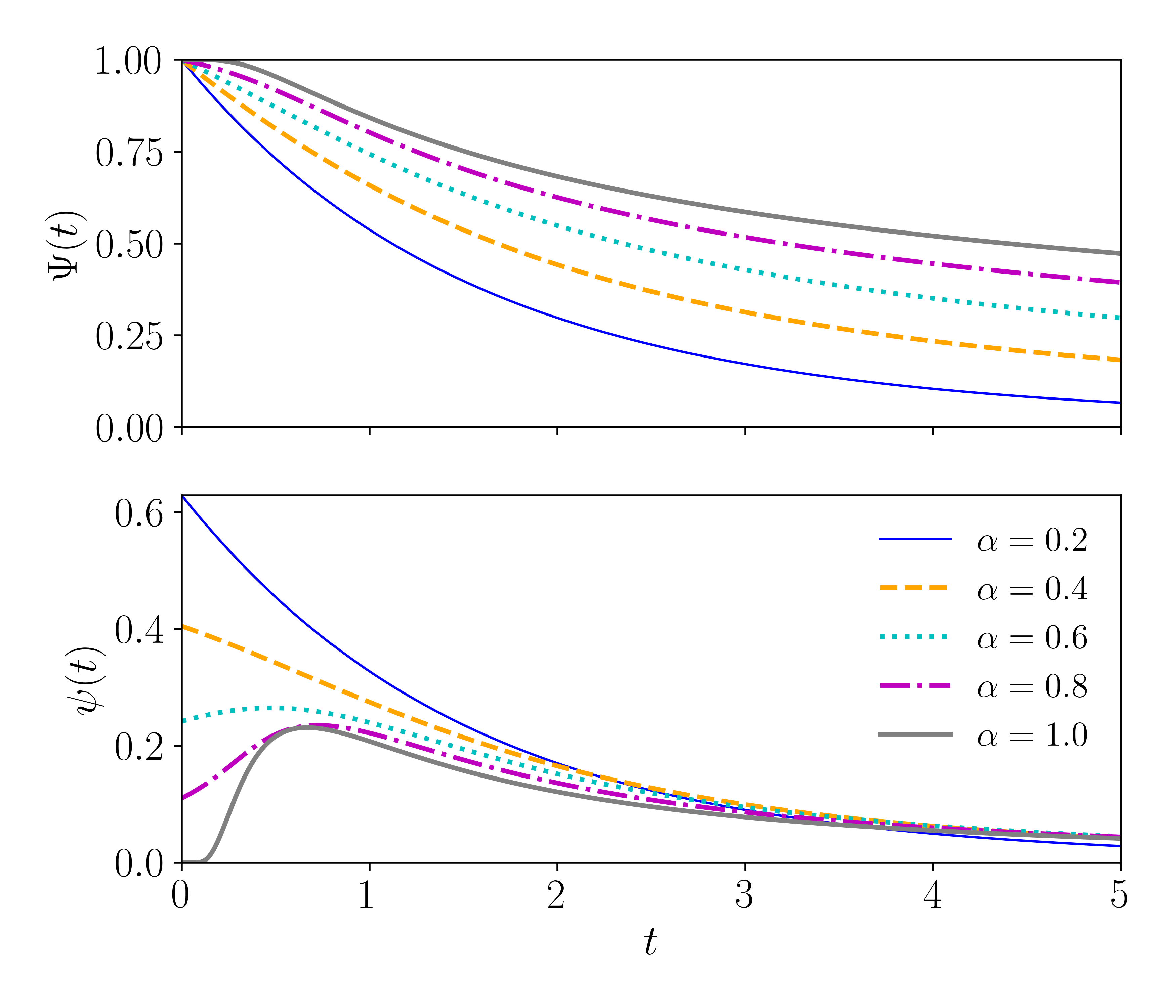}
	\caption{Plot of the survival function \eqref{UnbiasedH_sur} (top) and PDF \eqref{UnbiasedH_PDF} (bottom) of the FPT for the one-sided absorbing diffusion process on the semi-infinite interval approximated with numerical integration. Here, $D_\alpha = 1$, $x_0 = 2$, $0 \leq t \leq 5$ for $\alpha = 0.2$, $0.4$, $0.6$, $0.8$ and $1.0$.}
	\label{fig:UnbiasedH_FPT}
\end{figure}

When the target region is only on one side of the domain, the governing equation is \eqref{eq:FracDiffEq} but now the boundary conditions become $\rho(0,t)= \lim_{L \rightarrow \infty} \rho(L,t)=0$.
By taking the limit as $L \rightarrow \infty$ such that \eqref{Unbiased2s_rho} becomes an infinite sum of Riemann integrals with $\omega = n/L$, we obtain
\begin{equation}\label{UnbiasedH_rho}
	\rho(x,t) = 2\int_0^\infty e^{-D_\alpha (\pi \omega)^{2\alpha} t}  \sin \left( \pi \omega x_0\right)  \sin \left( \pi \omega x \right) \ d\omega,
\end{equation}
for $x \in [0,\infty)$.
Taking the same limit of \eqref{Unbiased2s_Sur}, the survival function becomes
\begin{equation}\label{UnbiasedH_sur}
	\Psi(t) = \dfrac{2}{\pi} \int_0^\infty \omega^{-1} e^{-D_\alpha (\pi \omega)^{2\alpha} t} \sin \left(\pi \omega x_0\right) \ d\omega,
\end{equation}
with PDF
\begin{equation}\label{UnbiasedH_PDF}
	\psi(t) = 2D_\alpha \int_0^\infty (\pi \omega)^{2\alpha-1} e^{-D_\alpha (\pi \omega)^{2\alpha} t}  \sin \left(\pi \omega x_0\right) \ d\omega.
\end{equation}
The survival function and PDF for the FPT are shown in Figure \ref{fig:UnbiasedH_FPT}.
For $\alpha = 1$, the survival function \eqref{UnbiasedH_sur} for the compounded random walk reduces to $\Psi(t) = \mathrm{erf} (x_0/\sqrt{4D_1t})$, which is consistent with the survival function for diffusion \cite{redner2001guide}. 

\begin{figure}[h!]
	\includegraphics[width=\linewidth]{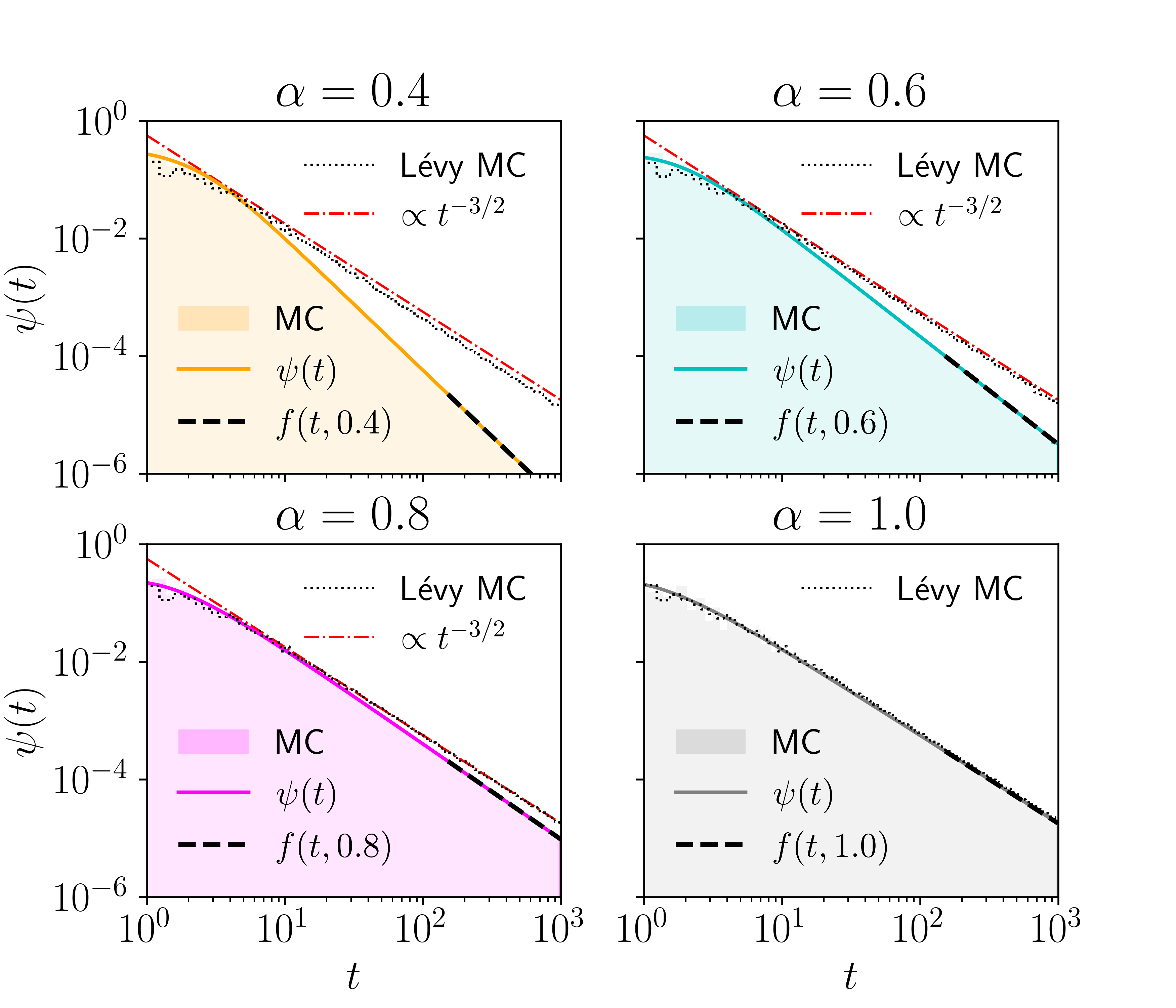}
	\caption{Log-log plot for the PDF \eqref{UnbiasedH_PDF} (solid lines) and its asymptotic behavior $f(t,\alpha)$ \eqref{asym} (dashed black lines) of the FPT for the one-sided absorbing diffusion process on the semi-infinite interval for $\alpha = 0.4$, $0.6$, $0.8$ and $1.0$ starting at $x_0 = 2$. The corresponding plot for the FPT Monte Carlo results of the underlying compounded random walk \eqref{eq:embedding} (solid filled bars) and L\'evy flight (dotted lines) for $10^6$ particles. Here, $\Delta t = 0.125$ and $\Delta x$ was determined by setting $D_\alpha = 1$ and using \eqref{eq:D_alpha} for $\alpha = 0.4$, $0.6$, $0.8$ and $1.0$, respectively. To compare the FPT PDF of L\'evy flights to the scaling $\propto t^{-3/2}$ we have included the red dotted-dashed lines in each subplot except for $\alpha = 1$, since $f(t,1.0) \propto t^{-3/2}$.}
	\label{fig:UnbiasedH_MC_quad}
\end{figure}

The PDF in \eqref{UnbiasedH_PDF} can be evaluated as a Fox H-function by identifying 
$$
e^{-z}= H_{0,1}^{1,0}\left[z\left|\begin{array}{c}
-\cr
(0,1)\cr
\end{array}
\right.\right]
$$
and using (2.49) from \cite{mathai2009h} to evaluate the Fourier sine integral. This results in
$$
\psi(t)=
\frac{2^{2\alpha}D_\alpha}{x_0^{2\alpha}\sqrt{\pi}}
 H_{2,1}^{1,1}\left[\frac{2^{2\alpha}D_\alpha t}{x_0^{2\alpha}}\left|\begin{array}{l}
(\frac{1}{2}-\alpha,\alpha),(1-\alpha,\alpha)\cr
(0,1)\cr
\end{array}
\right.\right],
$$
which has the long time power law scaling
\begin{equation}\label{eq:psiscaling}
\psi(t)\sim \frac{\Gamma(\frac{1}{2\alpha})x_0}{2\alpha^2 \pi D_\alpha^{\frac{1}{2\alpha}}} t^{-1-\frac{1}{2\alpha}}.
\end{equation}
The power-law scaling of the survival function \eqref{UnbiasedH_sur} for large time can also be found using the inequality
\begin{equation}
    \int_0^{\pi \omega x_0} (1-u) \ du \leq \int_0^{\pi \omega x_0}\cos(u) \ du \leq \int_0^{\pi\omega x_0} \ du,
    \label{eq:cosineIntegralInequality}
\end{equation}
where $\omega, x_0 >0$. 
Evaluating \eqref{eq:cosineIntegralInequality} gives
\begin{equation}
    \pi\omega x_0 - \frac{(\pi \omega x_0 )^2}{2} \leq \sin(\pi \omega x_0) \leq \pi \omega x_0.
    \label{eq:SineInequality}
\end{equation}
Using \eqref{eq:SineInequality} in \eqref{UnbiasedH_sur}, we obtain
\begin{equation}\label{asym}
	f(t,\alpha) - \dfrac{x_0^2 \Gamma (1/\alpha)}{2\alpha \pi D_\alpha^{\frac{1}{\alpha}} t^{\frac{1}{\alpha}}} \leq \Psi(t) \leq f(t,\alpha),
\end{equation}
where 
\begin{equation}
    f(t,\alpha) = \dfrac{x_0 \Gamma (1/(2\alpha))}{\alpha \pi D_\alpha^{\frac{1}{2\alpha}} }t^{-\frac{1}{2\alpha}}.
\end{equation}
From this we observe that the PDF of the FPT for the compound random walk on the semi-infinite line follows a power law distribution \eqref{eq:psiscaling}, which is different to those of the L\'evy flight.
Figure \ref{fig:UnbiasedH_MC_quad} shows the agreement between the power-law tail of the PDF $\psi(t) \sim t^{-1/(2\alpha)-1}$ obtained from \eqref{asym}, the analytical expression \eqref{UnbiasedH_PDF} and stochastic simulations of the compounded random walk.
Figure \ref{fig:UnbiasedH_MC_quad} also shows the PDF of L\'evy flights from stochastic simulations and the predicted $\sim t^{-3/2}$ asymptotic tail \cite{zumofen1995absorbing,chechkin2003first,palyulin2019first}.

The scaling that we obtained for the FPT density for the compounded walk is exactly the same scaling that would be obtained using the method of images for \eqref{eq:FracDiffEq} on an infinite domain, see also \cite{chechkin2003first}.
Note however, that this scaling is different to the scaling of FPT density for L\'evy flights due to the lack of continuity in trajectories required for the method of images \cite{chechkin2003first}.
For the compounded random walk, the particle may be absorbed along its path from the initial to the final location during each compounded step.
Therefore, its FPT will always be faster than its L\'evy flight counterpart, which is equivalent to a jump from the initial to the final location of the compounded step neglecting the path the particle may have taken \cite{palyulin2019first,Angstmann2025}.
This leads to a substantial difference in the FPT PDF as shown in Figure \ref{fig:UnbiasedH_MC_quad}.
For $\alpha = 1$, the compounded random walk and the L\'evy flight share the same FPT PDF since both are diffusive with power-law tail $\sim t^{-3/2}$.
As $\alpha$ decreases, Figure \ref{fig:UnbiasedH_MC_quad} shows the L\'evy flight FPT PDF ($\propto t^{-3/2}$ in accordance to Sparre-Andersen \cite{zumofen1995absorbing,chechkin2003first,palyulin2019first}) becomes increasingly different to the compounded random walk FPT PDF ($\propto t^{-1/(2\alpha)-1}$).
If the FPT PDF is calculated by integrating the governing equation \eqref{eq:FracDiffEq} or \eqref{specfracFP} over the domain, we obtain the same result as the compounded random walk.
This is not true for the L\'evy flight and suggests that the compounded random walk is a more accurate microscopic model for the superdiffusion equation \eqref{eq:FracDiffEq}.
In previous work, the PDF for the L\'evy flight differs when considering first arrival times at a point and first passage times when traversing a barrier \cite{chechkin2003first}.
This is disturbing physically as it implies that the underlying model could pass by a barrier without hitting any point located before the barrier, even in one dimension.

\begin{figure}[h!]
	\includegraphics[width=\linewidth]{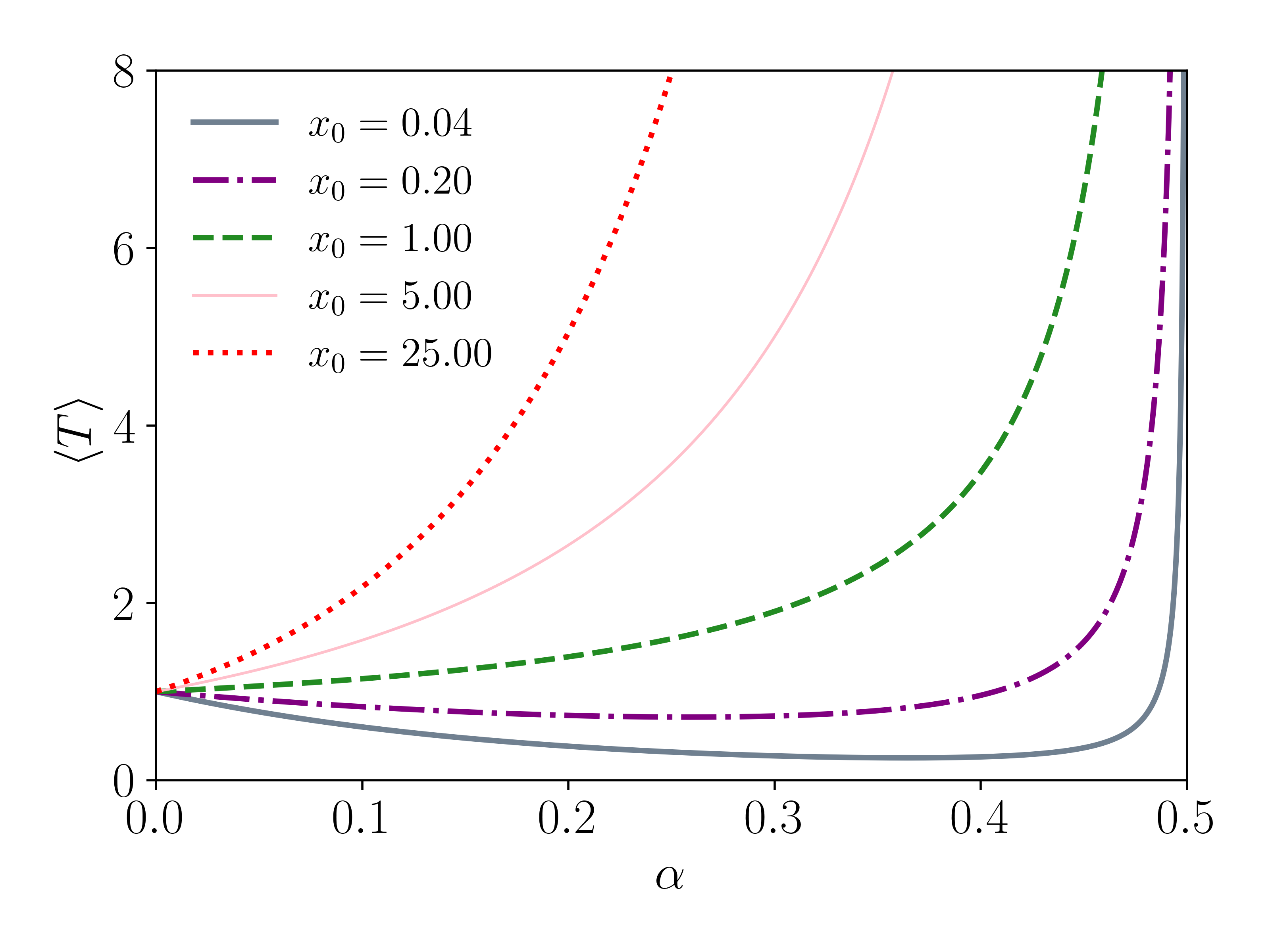}
	\caption{The expected FPT for the one-sided absorbing diffusion process on the semi-infinite interval \eqref{UnbiasedH_exp}. Here, $D_\alpha = 1$ and $0.0 < \alpha < 0.5$ for $x_0 = 0.04$, $0.20$, $1.00$, $5.00$ and $25.00$.}
	\label{fig:UnbiasedH_exp}
\end{figure}

Using \eqref{UnbiasedH_PDF}, we obtain
\begin{equation}\label{UnbiasedH_exp}
	\begin{split}
		\langle T \rangle &= \dfrac{2}{D_\alpha} \int_0^\infty (\pi \omega)^{-2\alpha-1} \sin \left(\pi \omega x_0\right) \ d\omega\\
		&= \dfrac{2x_0^{2\alpha}}{D_\alpha \pi} \cos \left( \dfrac{(2\alpha+1)\pi}{2}\right) \Gamma(-2\alpha)\\
		&= \dfrac{x_0^{2\alpha}}{\alpha D_\alpha \pi} \sin \left( \alpha \pi \right) \Gamma(1-2\alpha),\\
	\end{split}
\end{equation}
for $0 < \alpha < \frac{1}{2}$. 
The existence of such an expression indicates that a finite mean first-passage time exists for $0 < \alpha < \frac{1}{2}$.
However, for $\alpha \geq \frac{1}{2}$, the expected FPT is divergent due to the behavior of the PDF power-law tail.
On the other hand, the expected FPT of the L\'evy flight on the half-line is divergent for all $0 < \alpha \leq 1$.
This is because the PDF of the FPT for the L\'evy flight follows $\psi(t) \sim t^{-3/2}$ \cite{zumofen1995absorbing,chechkin2003first,palyulin2019first} as seen in Figure \ref{fig:UnbiasedH_MC_quad}.
For the compounded random walk, Figure \ref{fig:UnbiasedH_exp} shows how the expectation varies for $0<\alpha<\frac{1}{2}$ for various initial positions. 
It can be seen in \eqref{UnbiasedH_exp} that for fixed $D_\alpha$ and small $x_0$, a non-trivial choice of $\alpha$ which optimizes the expected FPT exists. 

As mentioned before, these FPT properties differ between the compounded random walk and the L\'evy flight due to the compounded random walk being able to interact with the local environment throughout its entire path. 
A jump that results in the particle moving from an initial position to a final position both on the semi-infinite line will never cross the origin for a L\'evy flight but the intermittent trajectory may have been absorbed for the compounded random walk.
This detail is highlighted by the difference between first arrival time and FPT for L\'evy flights, due to the ability to jump over targets \cite{chechkin2003first,palyulin2019first,gomez2024first,tzou2025green}, but their equivalence for compounded random walks.
This is due to the continuous sampling of the intermittent trajectory, meaning any trajectory which passes over the target must also have hit that target.
It may seem that the underlying compounded random walk, despite having symmetrical steps, contradicts the Sparre-Anderson Theorem, which dictates that $\psi (t) \sim t^{-3/2}$ \cite{redner2001guide}. 
Nonetheless, this has not occurred.
The underlying discrete steps are symmetric and i.i.d., so that the Sparre-Andersen theorem applies at the level of the step number $n$ giving $P(n^*<n) \sim n^{-1/2}$.
However, the embedding to continuous time via Sibuya-distributed compounding yields the time scaling, $P(T_{n^*}<t) \sim t^{-1/(2\alpha)}$.

\section{Stochastic simulation method}
The simulation of each step of the compounded random walk would have a divergent expected simulation time due to the divergent mean in each Sibuya random variable. Instead, we consider the location of the particle after each time step and find the conditional probability that the particle was absorbed during the intermittent steps. The simulation method for the Monte Carlo scheme is performed as follows:
\begin{enumerate}
    \item Choose the value of $\Delta t$ and calculate the value of $\Delta x$ via the relation \eqref{eq:D_alpha}
    \item Initialize particle at some position $X = x_0$ and set initial time as $t = 0$
    \item Draw a Sibuya random variable $K$ \cite{Hofert2011,leonenko2025sibuya}
    \item Draw a binomial random variable $Y$ for the number of steps to the right so that the number of steps taken to the left is $K-Y$
    \item Calculate the probability $P$, that the particle will be stopped in the next interval of $\Delta t$ time with the following rules:
    \begin{enumerate}[label=(\roman*)]
        \item If $X \leq 0$, then $P=1$
        \item If $X > 0$ and $X + Y\Delta x \leq K\Delta x$, then we have $$P = \frac{\Gamma(Y+1)\Gamma(K-Y+1)}{\Gamma(\lceil X/\Delta x \rceil +Y+1)\Gamma(K-\lceil X/\Delta x \rceil -Y+1)}$$
        \item Otherwise, $P = 0$
    \end{enumerate}
    \item Update $t \leftarrow t + \delta t$
    \item Draw a Bernoulli random variable with parameter $P$ to stop the process
    \item If not stopped, then given $X = x_j$, update $X \leftarrow X + (2Y-K)\Delta x$
    \item Repeat steps (3)-(8) until the particle is stopped or $t \geq T$, where $T$ is the simulation end time
\end{enumerate}
The time at which the particle is stopped is the FPT. The FPT for the Monte Carlo compounded random walk (solid filled bars) in Figure \ref{fig:UnbiasedH_MC_quad} were generated using this method.

\section{Conclusion}
We conclude that the space-fractional spectral Fokker-Planck equation \eqref{specfracFP} arising from a compounded random walk is amenable to first-passage time problems.
Here we have embedded a discrete-time random walk in real time and taken a diffusion limit to obtain the governing equation \eqref{specfracFP}, the same equation can be derived using a CTRW framework \cite{Angstmann2025}.
The underlying compounded process directly corresponds to the physical path which a particle traverses, thus permitting local effects from a potential to properly influence its trajectory.
This property also makes the first hitting or arrival time equivalent to the FPT, unlike the L\'evy flight.
Furthermore, the spectral nature of the governing equation allows us to obtain analytical solutions for the FPT distribution and its properties given different constraints.
These analytical solutions have a natural connection with their diffusion FPT counterparts making the superdiffusive results physically interpretable. 

We examined the difference between the two models in the semi-infinite domain and found the compounded random walk has a finite mean FPT for $0 < \alpha < 1/2$, as opposed to the divergent mean FPT for L\'evy flights \cite{palyulin2019first}.
Surprisingly, we find that there is a choice of $\alpha$ dependent on the initial position $x_0$ and $D_{\alpha}$ that minimizes the mean FPT for the compounded random walk.
Thus, the compounded random walk model introduced in this paper provides an alternative underlying microscopic model that shares the governing equation and pseudo-MSD properties with L\'evy flights but also retaining the spectral and asymptotic properties with the fractional Fokker-Planck equation.
Finally, we detail an efficient Monte Carlo method for stochastic simulation of these compounded random walks.
This makes the spectral fractional Fokker-Planck equation and its underlying compounded random walk an ideal tool for modeling across a range of physical systems.


\bibliography{mainbib}

\end{document}